\documentclass[10pt,twocolumn,pra,aps,superscriptaddress, floatfix]{revtex4-2}
\usepackage[english]{babel}
\addto\captionsenglish{}
\usepackage{amsmath}
\usepackage{graphicx}
\usepackage{soul}
\usepackage[colorlinks=true, allcolors=blue]{hyperref}
\usepackage{mathrsfs}

\usepackage{csquotes}
\usepackage{xcolor}
\usepackage{soul}
\usepackage{orcidlink}
\usepackage{ragged2e}

\begin{document}
\title{
Quantum Otto machine with $q$-deformed P\"oschl-Teller oscillator
}
\author{Collins O. Edet}
\email{collinsokonedet@gmail.com}
\affiliation{Faculty of Electronic and Engineering Technology, Universiti Malaysia Perlis, 02600 Arau, Perlis, Malaysia}
\affiliation{Department of Physics, University of Cross River State, Calabar, Nigeria}

\author{Norshamsuri Ali}
\affiliation{Faculty of Electronic and Engineering Technology, Universiti Malaysia Perlis, 02600 Arau, Perlis, Malaysia}
\affiliation{Centre of Excellence Advanced Communication Engineering (ACE), Universiti Malaysia Perlis, 02600 Arau, Perlis, Malaysia}

\author{Rosdisham Endut}
\affiliation{Faculty of Intelligent Computing, Universiti Malaysia Perlis, 02600 Arau, Perlis, Malaysia}
\affiliation{Centre of Excellence Advanced Communication Engineering (ACE), Universiti Malaysia Perlis, 02600 Arau, Perlis, Malaysia}

\author{O. Abah}
\email{obinna.abah@newcastle.ac.uk}
\affiliation{School of Mathematics, Statistics and Physics, Newcastle University, Newcastle upon Tyne NE1 7RU, United Kingdom}

\begin{abstract}
We study the impact of the potential parameters of the q-deformed modified Pöschl–Teller potential on the thermodynamic performance of a quantum Otto cycle, where the $q$-deformed modified Pöschl–Teller potential serves as the working substance. Analytical expressions for the energy spectrum and wave functions are derived, enabling a systematic investigation of heat exchange, work output, efficiency, and coefficient of performance. We show that $q$-deformation modifies the energy spectrum and creates distinct performance regions in the ($q$, $\Delta$) parameter space. Low ($\Delta$) and high ($q$) favour optimal heat engine efficiency, whereas high ($\Delta$) and low ($q$) improve refrigerator performance. The heat engine efficiency peaks in the low-($\Delta$), high-(\texorpdfstring{$q$}{q}) regime. These results highlight the q-deformed modified Pöschl–Teller potential as a versatile and tunable platform for exploring potential parameter-driven effects in quantum thermal machines.
\end{abstract}
\maketitle

\section{Introduction}
Heat engines and refrigerators have been fundamental to the development of modern society, from the Industrial Revolution to the operation of contemporary technologies. While heat engines produce useful work by absorbing heat from a hot reservoir, refrigerators consume work to extract heat from a cold reservoir. They also play a crucial role in elucidating the thermodynamic behavior of a broad class of physical systems \cite{kondepudi2014modern}. In 1959, the seminal work of Scovil and Schulz-DuBois on the heat engine \citet{scovil1959three} and later on the refrigerator \cite{Geusic67} started a paradigm shift by extending the thermodynamic framework of the thermal machines to the quantum domain, thereby opening the way for the application of thermodynamic concepts at the quantum scale. Following the advancement in nanofabrication techniques and coherent atomic control, quantum thermal machines have emerged as a promising framework for investigating the thermodynamic properties of quantum systems and the fundamental role of information in quantum thermodynamics \cite{vinjanampathy2016quantum, goold2016role, binder2018thermodynamics, deffner2019quantum, bhattacharjee2021quantum,myers2022quantum, potts2024quantum}. Extensive research has focused on examining the influence of various quantum features on the performance of thermal machines, including quantum correlations \cite{barrios2017role}, many-body effects \cite{mukherjee2021many, rolandi2023collective}, quantum uncertainty \cite{kalaee2021violating}, degeneracy \cite{pena2017magnetic, alvarado2018quantum}, endoreversible cycles \cite{smith2020endoreversible, Myers2021Thermodynamics}, finite-time cycles \cite{feldmann2012short,Zheng2016Occurrence, Cavina2017Slow, raja2021finite}, energy optimisation \cite{Singh2022Unified}, shortcuts-to-adiabaticity \cite{del2013shortcuts, campbell_trade-off_2017, abah_performance_2018}, efficiency and power statistics \cite{denzler2021efficiency,denzler2021power, denzler2024nonequilibrium}, coherence \cite{scully2003extracting, feldmann2012short, dorfman2013photosynthetic, hardal2015superradiant, uzdin2016coherence, watanabe2017quantum, dann2020quantum, hammam2021optimizing}, as well as comparative analyses between classical and quantum thermal machines \cite{quan2007quantum,abah2012single, gardas2015thermodynamic, friedenberger2017quantum}.

Since the rise of quantum thermodynamics as a framework for understanding nonequilibrium quantum systems, intense efforts have been intense efforts directed toward the realization of quantum thermal machines using different working media \cite{quan2007quantum, rossnagel2016single, gelbwaser2018single}. The proposed implementation platforms range from magnetic systems \cite{pena2015magnetostrain}, atomic clouds \cite{niedenzu2019quantized}, transmon qubits \cite{cherubim2019non}, harmonically confined single ions \cite{ abah2012single}, optomechanical systems \cite{zhang2014quantum, dechant2015all}, and quantum dots \cite{pena2019magnetic,pena2020quasistatic}. Moreover, experimental realizations have been demonstrated in atomic collisions \cite{bouton2021quantum}, two-level ions \cite{van2020single}, nuclear magnetic resonance \cite{peterson2019experimental}, an ensemble of nitrogen vacancy center \cite{klatzow2019experimental}, and single-ion \cite{van2020single}. 

Furthermore, the efficiency of a quantum Otto engine has been shown to be enhanced by using a working medium whose energy levels scale inhomogeneously \cite{gelbwaser2018single}. The P\"oschel-Teller potential, which has unequal energy level spacing and captures the nonlinear physics of diatomic molecules \cite{edet2020thermal} and quantum dots \cite{Yildirim2005}, has been employed as a working medium of a quantum engine cycle \cite{oladimeji2019,Abasabadi2023}. A recent study showed that the quantum statistical deformation parameter $q$ can influence the efficiency and work of the quantum Otto cycle \cite{Ozaydin2023}.


Over the past five decades, $q-$deformed algebra has garnered significant attention within the quantum science community \cite{bayindir2021self,
chung2024q,
mousavigharalari2025q,
demirbilek2025analytical,
altintas2020q}. It was initially introduced as generalizations of the Weyl-Heisenberg algebra \cite{arik1975operator, arik1976hilbert}. In this framework, $q$-deformation has been extensively studied in a wide range of research areas, including nuclear and atomic physics \cite{bonatsos1992generalized}, thermodynamics \cite{naseri2022non}, statistical physics and quantum information \cite{altintas2014constructing}, open quantum systems \cite{naseri2022non},  optomechanical systems \cite{kundu2022transparency}, and many-body interactions in nuclear systems \cite{sviratcheva2004physical}. Considerable effort has been devoted to uncovering physical realizations of the deformation parameter $q$, with notable examples arising in the quantum Yang-Baxter equation \cite{ma1993yang}, the deformed Jaynes-Cummings model \cite{chaichian1990quantum}, the quantum phase problem \cite{ellinas1992quantum, hakioglu1998admissible, hakioglu1998finite}, the relativistic $q$-oscillator \cite{mir1991suq, arik1992q}, the Morse oscillator \cite{cooper1995q}, and the Kepler problem \cite{dayi1995slq}. The $q$-deformation parameter has been employed to derive generalized uncertainty relations, information-theoretic measures \cite{van1984generalized}, Tsallis entropy, and other generalized formulations of relative entropy \cite{naudts2011generalised, borges1998family, landsberg1998distributions}. In recent time, the deformed algebra has been used to study the relationship between the efficiency of quantum heat engine and the degree of the non-Markovianity \cite{naseri2022non}.

In this paper, we examine the performance of a quantum Otto machine with a working medium modelled by the q-deformed modified P\"oschl-Teller potential. We analyze when the thermal cycle performs as a heat engine and, then as a refrigerator. In section \ref{model} we provide some necessary background on the q-deformed modified Pöschl-Teller potential, a description of the model alongside the energy and wave function expressions. In section \ref{quantumotto}, we present the quantum Otto cycle with the q-deformed modified Pösch-Teller potential and show the regimes where the system behaves either like a refrigerator or heat engine in the parameter regimes considered. Finally, in section \ref{conclusion} we conclude with a discussion of possible experimental systems in which such a thermal cycle could be implemented.


\begin{figure*}
    \centering
    \includegraphics[width=0.95\linewidth]{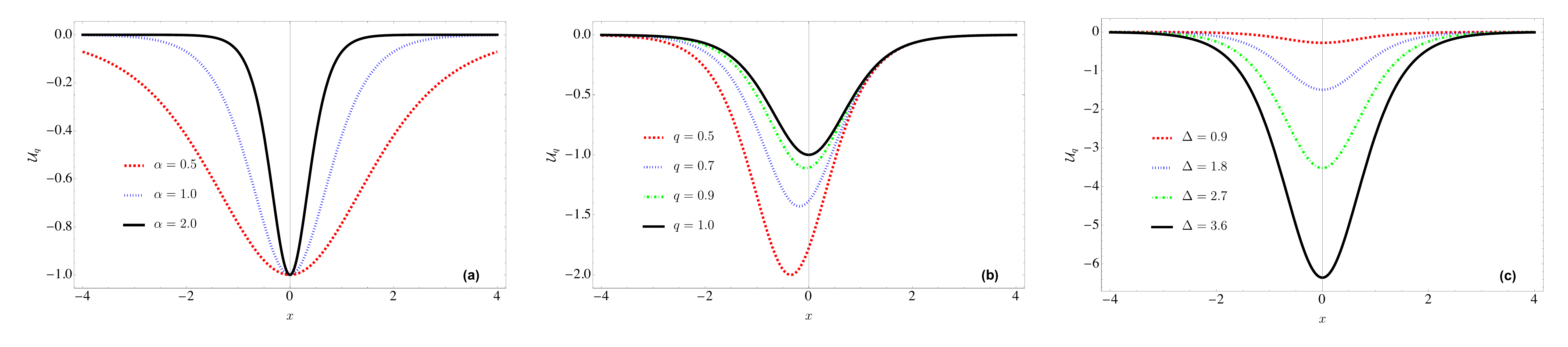}
    \caption{\justifying The $q$-deformed modified Pöschl–Teller potential as a function of $x$ for different: \textbf{(a)}  $\alpha = 0.5$ (dotted red line), $\alpha = 1$ (dot blue line), and $\alpha = 2.0$ (solid black line) \textbf{(b)} $q = 0.5$ (dotted red line), $q = 0.7$ (dot blue line),  $q = 0.9$ (dot green line)and $q = 1.0$ (solid black line) (standard Pösch-Teller potential). \textbf{c} $\Delta = 0.9$ (dotted red line), $\Delta = 1.8$ (dotted blue line), $\Delta=2.7$ (dotted green), and $\Delta = 3.6$ (solid black line). Other parameters used: $q=1$, $\Delta\!=\!1.5$, and $\alpha\!=\!1$. }
    \label{figpotential}
\end{figure*}

\section{\texorpdfstring{$q$}{q}-deformed P\"oschl-Teller potential}\label{model}
\begin{figure*}
    \centering
    \includegraphics[width=1\linewidth]{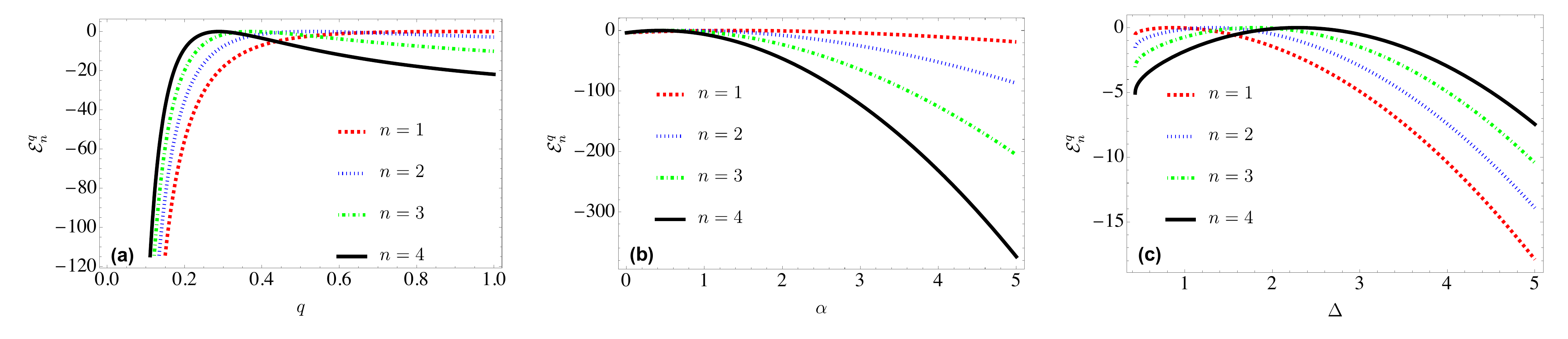}
    \caption{ \justifying Energy $\mathcal{E}_{n}^{q}$ of the $q$-deformed  P\"oschl–Teller potential as a function of \textbf{(a)} $q$  \textbf{(b)} $\alpha$ and \textbf{(c)} $\Delta$, for different energy level $n$. Here, $n=1$ (dotted red line), $n = 2$ (dot blue line), $n = 3$(dot green line) and $n = 4$ (solid black line). Other parameters used: $q=1$, $\Delta\!=\!2$, and $\alpha\!=\!1.5$ (a) and $\alpha\!=\!0.5$(b).}
    \label{figenergy}
\end{figure*}
The Hamiltonian of a quantum system with the $q$-deformed P\"oschl–Teller potential is \cite{grosche2005path}
\begin{equation}
 \hat{\mathcal{H}}= \frac{\hat{p}^2}{2 \mu}+ \mathcal{U}_q(x) = \frac{\hat{p}^2}{2 \mu}   - \left( \frac{\hbar^{2}}{2 \mu \cosh_{q}^{2}(\alpha x)} \left(\Delta^{2}-\frac{1}{4}\right) \right),   
\end{equation} 
where the first term is the kinetic energy operator  with $p\!=\!-i\hbar d/dx$ defined as the momentum operator, and $d/dx$  is the differential operator in $\text{1D}$, and $\mu$ is the reduced mass. The second term represents the potential energy function $\mathcal{U}_{q}(x)$, with $\Delta$ being the parameter that controls the depth of the potential well (equivalent to the dissociation energy in diatomic molecular systems), the parameter $\alpha$  controls the width of the well, $x \in \Re$ is the internuclear separation, and $q$ is the deformation parameter typically within the range $0<q<1$. 
It is based on the q-deformation of the usual hyperbolic potential; $\cosh_q y = \frac{1}{2} \left(e^{y} + q e^{-y}\right)$ and $\sinh_q y = \frac{1}{2} \left(e^{y} - q e^{-y}\right)$, where we assume without loss of generality that $q>0$. This potential belongs to the class of shape invariant potentials derived from supersymmetric quantum mechanics \cite{fischer1993path, grosche1998handbook}. This was  first introduced by \citet{arai1991exactly} and was also discussed by \citet{kalnins1997superintegrability} and Lemieux and Bose \cite{lemieux1969construction}  in the context of general solutions of the hypergeometric equation. 
Figure \ref{figpotential} illustrates the behavior of the  $q$-deformed Pöschl–Teller potential. In Fig \ref{figpotential}(a), we illustrate the shape of the $q$-deformed P\"oschl--Teller potential $\mathcal{U}_q(x)$ as a function of the spatial coordinate $x$ for different values of
the  parameter $\alpha$. Here, the potential exhibits a symmetric well that is centered at $x=0$ for all values of $q$. When $\alpha$ decreases, the width of the potential well increases. For large values of $\alpha$, the potential becomes narrow. Fig \ref{figpotential} (b) shows a similar behavior, but with a more pronounced dependence on the deformation parameter. As $q$ decreases from unity, the potential well becomes progressively shallower. In the limit $q\rightarrow 1$, the standard hyperbolic Pöschl–Teller potential is recovered. The deformation thus modifies the shape of the potential without altering its symmetry or asymptotic behavior. Physically, the $q$-parameter provides a mechanism for the tunability of the potential energy function.  Fig. \ref{figpotential}(c) shows the dependence of the potential on the parameter $\Delta$, thereby increasing $\Delta$ results in a substantial deepening of the potential well, while the spatial width remains largely unaffected. The minimum potential scales with $-\left(\Delta^{2}-\frac{1}{4}\right) $, which  emphasizes the direct role of $\Delta$ in controlling the depth of the well.

For completeness, the solutions of the Schrodinger equation for the model potential have been presented in Appendix \ref{solutionsposch}. The  non-relativistic $q$-deformed modified P\"oschl–Teller potential eigenvalues reads
\begin{equation} \label{energyeq}
\mathcal{E}_n^{q} = - \frac{\alpha^{2} \hbar^{2}}{2 \mu} \left[ \left( n + \frac{1}{2} \right) - \Tilde{w} 
\right]^{2}.
\end{equation} 
where $\Tilde{w}\!=\!\sqrt{\frac{1}{4} + \frac{\xi}{q}}$ and the corresponding wave function is presented as follows:
\begin{equation}
\Psi_n^{q} = N^{q} (1 - z^{2})^{\frac{\xi}{2}} \, {}_{2}F_{1} \left( -n, -n + 2\Tilde{w} , -n + \Tilde{w} + \frac{1}{2}; \frac{1 - z}{2} \right),
\end{equation}
where ${}_{2}F_{1} \left(-n, -n + 2w, -n + w + \frac{1}{2}; \frac{1 - z}{2} \right)$ are the hypergeometric polynomials of degree $n$. Figure~\ref{figenergy}(a) shows the variation of the bound-state energy $\mathcal{E}_{n}^{q}$ of the $q$-deformed P\"oschl--Teller potential as a function of the deformation parameter $q$ for different quantum numbers $n$. For all quantum numbers considered, the energies increase monotonically as $q$ approaches unity, becoming less negative. 
Higher energy states exhibit a stronger sensitivity to the deformation parameter, as evidenced by the increasing separation between the curves. Figure~\ref{figenergy}(b) shows that as $\alpha$ increases, the energy levels decrease rapidly. The effect is particularly pronounced for higher excited states, whose energies decrease more with increasing  $\alpha$. The dependence of the bound-state energies on the  parameter $\Delta$ is presented in Fig.~\ref{figenergy}(c). The energy eigenvalues decrease monotonically with an increasing value of $\Delta$.

The canonical partition function of the working substance with energy level $\mathcal{E}^{q}_{n}$ in Eq. (\ref{energyeq}) at temperature $T$ is given by; $Z = \sum_n \exp(-\mathcal{E}_{n}^{q} / k_B T)$, which explicitly reads:
\begin{equation}
Z   =  \frac{\sqrt{\pi } \left(-\text{Erfi}\left(\sqrt{\beta} \sqrt{\text{p}} \sigma \right)\right)}{2 \sqrt{\beta} \sqrt{\text{p}}},
\end{equation}
where $\beta\!=\!1/k_BT$, $k_B$ being the Boltzmann constant, $\text{p}\!=\!\frac{\alpha^2 \hbar ^2}{2 \mu }$, \text{and} $\sigma\!=\!\frac{1}{2}-\sqrt{\frac{\Delta^2-\frac{1}{4}}{\alpha^2 \text{q}^2}+\frac{1}{4}}$. In the thermal equilibrium state, the density matrix is given as; $\rho =\sum_n \ \exp\!\left(-\frac{\hat{\mathcal{H}}}{k_B T}\right)\!/Z \, |\psi_n \rangle\langle \psi_n| = \sum_n P_n\, |\psi_n \rangle\langle \psi_n|$, where $P_n = \exp(-\mathcal{E}_{n}^{q} / k_B\,T) / Z$  is the occupation probability of the $n^{th}$ eigenstate.


\section{Quantum Otto cycle} \label{quantumotto}
\begin{figure} 
    \includegraphics[width=0.95\linewidth]{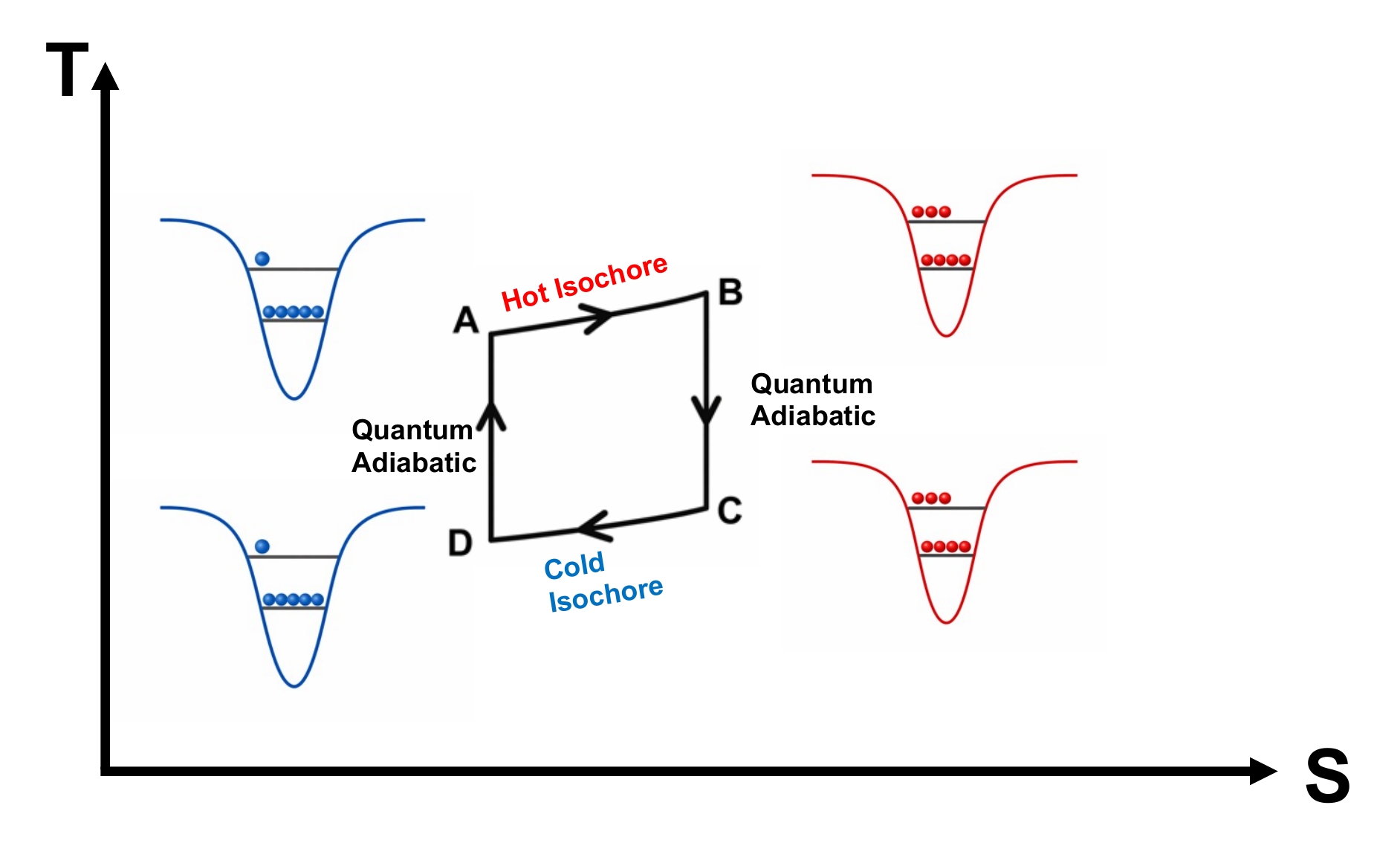}
    \caption{ \justifying The schematic of the Entropy-temperature (S-T) diagram of the quantum Otto cycle. The quantum thermal cycle comprises of two adiabatic strokes ($B\rightarrow C$ and $D\rightarrow A$) where it is decoupled from the thermal reservoirs, and two isochoric strokes ($A\rightarrow B$ and $C \rightarrow D$) where the engine is coupled to two thermal baths at temperatures $T_{h}$ and $T_c$, with $T_h > T_c$.}
   \label{Figschem}
\end{figure}

We present the quasi-static quantum Otto engine cycle, which operates in four strokes: two quantum isochoric processes and two quantum adiabatic processes (see Fig. \ref{Figschem}). In the first stroke, quantum isochoric heating process $(A \rightarrow B)$, the working substance ($q$-deformed modified hyperbolic P\"oschl-Teller potential) is coupled to the hot reservoir at temperature $T_h$. The system is transformed ($A \to B$) to a thermal state at point $B$ such that $\mathcal{E}_n^{q (B)}\!=\!\mathcal{E}_n^{q (A)}$ and population is changed to  $P_n^{q (B)}$. The heat absorbed by the working medium during the heating process ($A \to B$) is
\begin{align}\label{QhotOtto}
    \mathcal{Q}_{\text{h}}^{q} = \int^{B}_{A}\, \mathcal{E}_n^{q} d P_n^{q} =\,\sum_{n}\mathcal{E}_n^{q(h)}[P_n^{q (B)}-P_n^{q(A)}],
\end{align}
where $\mathcal{E}^{\text{h}}$ is the $n^{th}$  energy of the system from $A\to B$. The heat transferred from the environment to the system, $Q_h^{q}$, can be explicitly derived (see Appendix \ref{Ottoequations} for the expression). 
The second stroke is a adiabatic expansion $(B \rightarrow C)$ of the potential well width when it is totally isolated from the environment.
The system satisfy the conditions $P_n^{q(C)}\!=\!P_n^{q(B)}$ while the energy eigenvalues are adiabatically changed to $\mathcal{E}_n^{q(C)}$ without no heat is exchanged between the system and the environment. 
Then, the third stroke is a quantum isochoric process $(C \rightarrow D)$ in which the working substance is put in contact to the cold reservoir at temperature $T_c$. The system is isochorically cooled to point $D$ such that the population is modified to $P_n^{q(D)}$ at fixed $\mathcal{E}_n^{q(D)}\!=\!\mathcal{E}_n^{q(C)}$. The heat released during the cooling process is presented as follows,
\begin{align}\label{QcoldOtto}
    \mathcal{Q}_{\text{c}} = - \int^{D}_{C} \mathcal{E}_n^{q (c)}dP_n \, = \sum\limits_{n}\mathcal{E}_n^{q (c)}[P_n^{(D)}-P_n^{(B)}],
\end{align}
where $\mathcal{E}^{\text{q (c)}}$ is the $n^\text{th}$ energy of the cooling process of the system from $D\to C$. The exact expression of the heat transferred from the working substance to the cold reservoir is shown in Appendix \ref{Ottoequations}.
In the fourth stroke, the cycle is complete by performing a quantum adiabatic compression ($D \rightarrow A$) under the condition $P_n^{q(A)}\!=\!P_n^{q(D)}$.

The total work done for the quantum Otto machine, based on the first law of thermodynamics, is the sum of $\mathcal{Q}_{\text{h}}$ and $\mathcal{Q}_{\text{c}}$, reads \cite{quan2007quantum, Prakash2022}
\begin{equation}\label{WorkdoneOtto}
\mathcal{W}^{\text{q}}  =\sum\limits_{n} \left[\mathcal{E}^{q (h)}_{n}- \mathcal{E}^{\text{q (c)}}_{n}\right]\left[P_n^{q(B)}-P_n^{q(D)}\right].
\end{equation}
When the quantum Otto cycle is functioning as a heat engine, the efficiency of the $q$-deformed modified P\"oschl-Teller quantum heat engine is defined as \cite{quan2005quantum, kieu2006quantum}
\begin{equation}
    \eta^{q} = \frac{\mathcal{W}^{q}}{ \mathcal{Q}_h^{q}},
\end{equation}
where $\mathcal{W}^{q}$ denotes the work extracted per cycle and $\mathcal{Q}_h^{q}$ is the heat absorbed from the hot reservoir. On the other hand, when the quantum Otto cycle is functioning as a refrigerator, the main difference is the reversal of the cycle's direction. This means heat $\mathcal{Q}_{c}^{q} > 0$ is extracted from the cold reservoir, while heat $Q_{h}^{q} < 0$ is released to the hot reservoir by the working substance. 
The performance of the refrigerator is quantified by the coefficient of performance (COP), which is defined as the ratio of the heat extracted from the cold reservoir to the total work done in a complete cycle. Therefore, the refrigerator coefficient of performance is given by
\begin{equation}
    COP^{q} =\frac{\mathcal{Q}_{c}^{q}}{\mathcal{W}^{q}}.
\end{equation}


\begin{figure*}
    \includegraphics[width=0.89\linewidth]{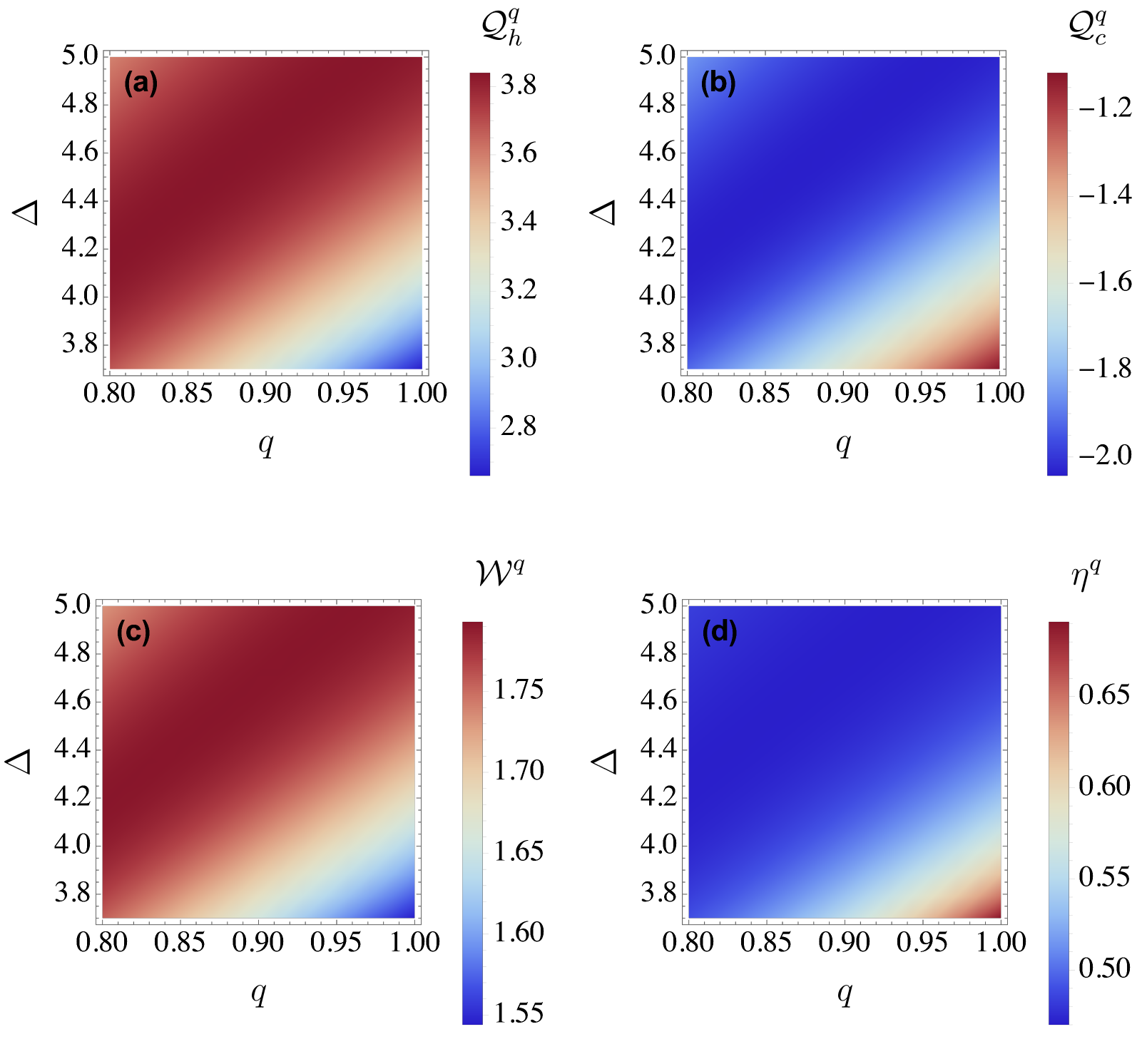} 
    \caption{\justifying Quantum Otto cycle as a heat engine. (\textbf{a}) Heat exchange with the hot reservoir,  $\mathcal{Q}_{\text{h}}^{q}$ (\textbf{b}) Heat exchange  with the cold reservoir,  $\mathcal{Q}_{\text{c}}^{q}$  (\textbf{c}) Total work output, $\mathcal{W}^{q}$  (\textbf{d}) Efficiency  $\eta^{q}$,  as function of the depth of the potential $\Delta (3.7<\Delta<5)$ and the deformation parameter $q (0.8<q<0.9)$. Parameters used: $\alpha_c\!=\!0.5$, $\alpha_h\!=\!1.118$, $T_h\!=\!5 $, and $T_c\!=\!1$. }
    \label{figengine}
\end{figure*}

\subsection{Quantum Otto heat engine analysis}
Figure \ref{figengine} shows the thermodynamics analysis of the quantum Otto cycle when it functions as a heat engine for different values of potential well depth $\Delta$ and deformation parameter $q$. It can be seen that $\mathcal{Q}_{h}^{q}$ increases with increasing $\Delta$, indicating an enhanced heat absorption for deeper potential wells; see \ref{figengine}(a). This behavior can be attributed to the larger energy level spacing associated with stronger confinement, which allows the working substance to absorb more energy during thermalization with the hot temperature reservoir. In contrast, the dependence on the deformation parameter $q$ is relatively weak. Figure \ref{figengine}(b) shows that the magnitude of heat exchanged during the cold isochoric process $\mathcal{Q}_h^q$ increases with $\Delta$, while the negative sign is attributed to  heat rejection from the system. As $q$ increases, $|\mathcal{Q}_{h}^{q}|$ decreases slightly, indicating reduced spectral compression and weaker thermal exchange during the cold isochore. Figure \ref{figengine}(c) displays positive work output throughout the considered parameter regime, confirming that the system operates as a quantum heat engine. As $\Delta$ increases from 3.0 to 5.0, the work output increases, as indicated by the color transition from blue (low work, $\sim 1.45$) to deep red (high work, $\sim 1.75$). The influence of the parameter $q$ is more subtle; for lower values of $\Delta$, increasing $q$ slightly decreases the work output, whereas for higher values of $\Delta$, the work remains consistently high and independent of $q$. In Fig. \ref{figengine}(d), we present the efficiency plot of the heat engine, which exhibits an almost inverse trend compared to the work output.
This behavior reflects the fact that increasing $\Delta$ raises the work output proportionally more than it raises the absorbed heat $\mathcal{Q}^{q}_h$, improving the ratio $\mathcal{W}^{q}/\mathcal{Q}^{q}_h$ despite the simultaneous increase in cold-bath heat rejection. Increasing q, by contrast, reduces both heat quantities but does so in a way that is less favorable to the work-to-heat ratio, thereby degrading efficiency. The obtained values remain below the Carnot limit corresponding to the chosen bath temperatures, confirming thermodynamic consistency.
The efficiency reaches its maximum value of approximately $0.6$ in the low-$\Delta$, high-$q$ regime and decreases gradually with increasing $\Delta$. The obtained values remain below the Carnot limit corresponding to the chosen bath temperatures, confirming thermodynamic consistency.
The parameter range shown in Fig. \ref{figengine}  was selected to ensure operation within the physically admissible positive-work regime and consistency with the Carnot bound.

\subsection{Quantum Otto refrigerator}
\begin{figure*}
    \includegraphics[width=0.89\linewidth]{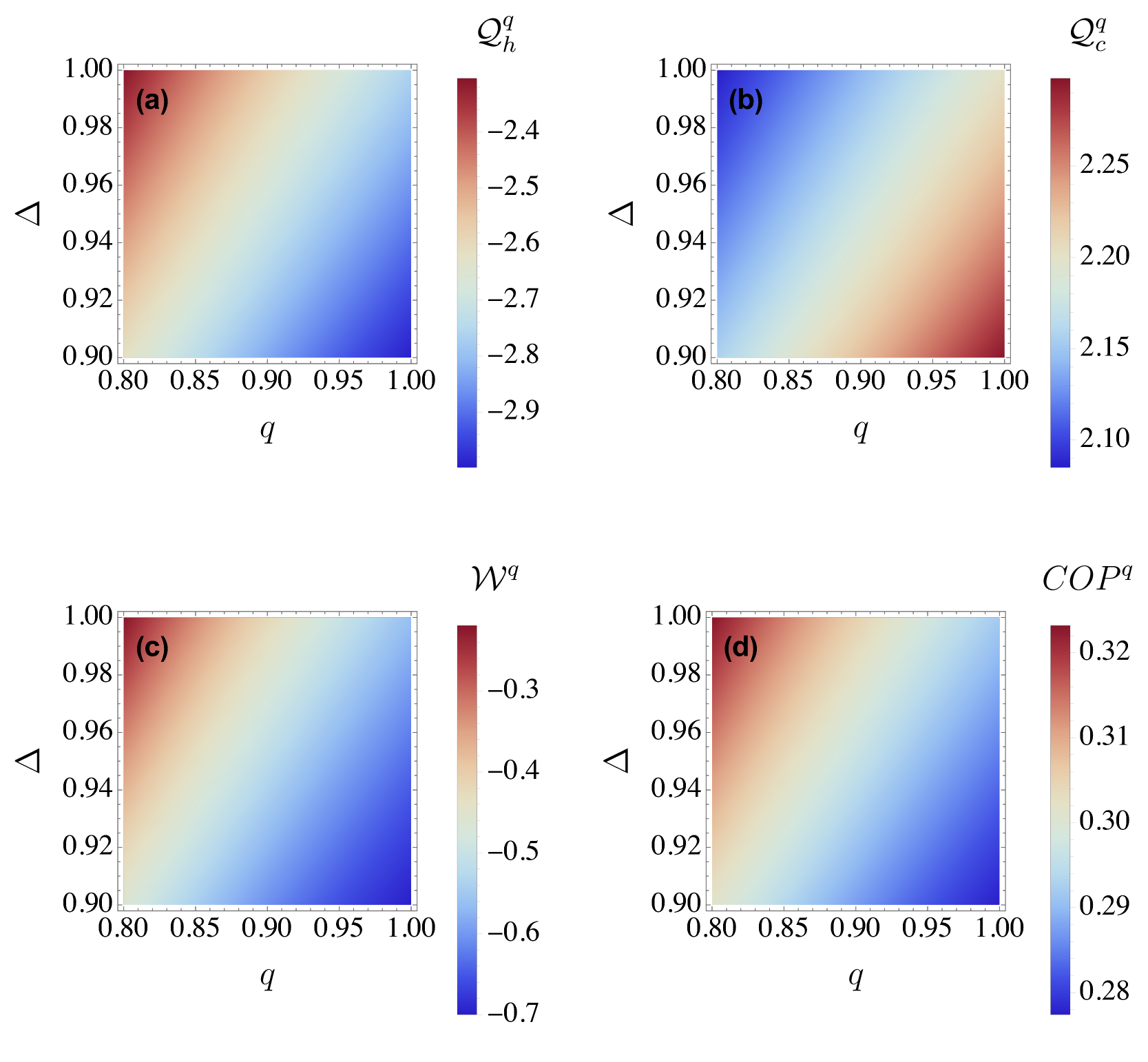}
    \caption{\justifying Quantum Otto refrigerator cycle (\textbf{a}) Heat exchange with the hot reservoir,  $\mathcal{Q}_{\text{h}}^{q}$ (\textbf{b}) Heat exchange  with the cold reservoir,  $\mathcal{Q}_{\text{c}}^{q}$  (\textbf{c}) Total work output, $\mathcal{W}^{q}$  (\textbf{d}) Coefficient of performance $COP^q$   as function of the potential well depth $\Delta (0.9<\Delta<1)$ and $q (0.80<q<1)$. Parameters used: $\alpha_c = 0.5 $, $\alpha_h = 1.118$, $T_h = 5 $, and $T_c=1$. }
    \label{figrefrigerator}
\end{figure*}

We now focus on the thermodynamic analysis of the quantum Otto cycle whose working substance is a $q$-deformed P\"oschl-Teller oscillator when it operates as a refrigerator. Figure \ref{figrefrigerator}(a) presents the heat exchanged with the hot reservoir in the quantum Otto refrigerator cycle as a function of $\Delta$ and $q$. Throughout the parameter space explored , $\mathcal{Q}_h < 0$, showing that the system releases heat to the hot temperature reservoir rather than absorbing it. The magnitude of the heat rejected by the system to the hot reservoir increases with increasing $q$, while it decreases as $\Delta$ increases. 
Figure \ref{figrefrigerator}(b) shows the heat exchanged with the cold temperature reservoir, $\mathcal{Q}_c$, which should be positive for the refrigeration operation. It can be seen that  $\mathcal{Q}_c$ increases monotonically with increasing $q$ and decreasing $\Delta$. The maximum values of $\mathcal{Q}_c$ are achieved in the regime of large  $q$ and small $\Delta$, whereas the minimum values occur for small $q$ and large $\Delta$. In Fig. \ref{figrefrigerator}(c) shows the work input $\mathcal{W}^q$ as a function of the parameters $q$ and $\Delta$. The work input increases with increasing $\Delta$ and decreases with increasing $q$. Consequently, the higher input work values are obtained in the regime of high values of $\Delta$ and small $q$, while the minimal work are obtained for small $\Delta$ and high $q$ values. Figure \ref{figrefrigerator}(d)  presents the coefficient of performance ($COP^q$) of the refrigerator as a function of the parameters $q$ and $\Delta$. The COP exhibits a smooth and monotonic dependence on the system parameters. Specifically, the $COP^q$ increases with increasing $\Delta$ and decreases with increasing $q$. This means that the most efficient refrigeration performance is achieved for large $\Delta$ and small $q$. Conversely, the lowest COP occurs for small $\Delta$ and large $q$.  
This occurs because larger $\Delta$ increases the accessible energy separation during the cooling stroke, enhancing heat extraction from the cold reservoir, while lower $q$ reduces the work input required per unit of cooling. Together, these effects make the high-$\Delta$, low-$q$ corner the most favorable operating point for refrigeration.

\section{Conclusions} \label{conclusion}
We have investigated the thermodynamic performance of a quantum Otto cycle using a $q$-deformed modified P\"oschl-Teller oscillator as the working substance. We have exploited the addition of $q$-deformation into the modified P\"oschl-Teller potential model to demonstrate that the quantum thermal device can perform in multiple operational regimes, including as quantum heat engine or a quantum refrigerator, depending on the chosen parameter range. We have presented the analytical expressions for the eigenenergies and wave functions of the potential model within a factorisation method, which allows the derivation a consistent thermodynamic description of the corresponding partition function. Then we analyzed the heat exchanged with the reservoirs, the work output/input, and the efficiency or coefficient of performance of the Otto cycle by varying the potential parameters, specifically the well width and deformation strength. Our results show that the potential parameters play a dominant role in controlling energy-level spacing and heat exchange and the thermodynamic behaviors. Moreover, the thermal device (engine or refrigerator) performance have a complementary dependence on the system parameters, indicating a trade-off between work output/input, cooling and performance that can be tuned through deformation and confinement. In summary, we have shown that $q$-deformation in a modified P\"oschl-Teller potential enables the emergence of new thermodynamic operational phases as well as enhances the key figures of merit of a quantum Otto thermal machine. Importantly, the proposed model is amenable to experimental realization, in a laser-cooled trapped ion as a microscopic heat machine \cite{gelbwaser2018single}.

\section*{Acknowledgments}
COE and NA  would also like to acknowledge the Universiti Malaysia Perlis (Funding No. 9004-00100 Special Research Grant-International Postdoctoral) for funding this project.

\appendix 
\section{ Solutions of Schrödinger equation with the q-deformed Pöschl-Teller potential} \label{solutionsposch}
The time independent Schrödinger equation of a $q$-deformed hyperbolic Pöschl-Teller potential reads \cite{grosche2005path, olusesi2025statistical}
\begin{equation}
\left[ \frac{d^{2}}{dx^{2}}- \frac{\left(\Delta^{2}-\frac{1}{4}\right)}{\cosh_{q}^{2}(\alpha x)} + \frac{2 \mu \mathcal{E}_{n}^{q}}{\hbar^{2}}
\right]\Psi(x) = 0,
\end{equation}
where $\Delta$ indicate the dissociation parameter, $\alpha$ represent the range parameter, $\mu$ denote the reduced mass. Applying the deformation hyperbolic functions \cite{arai2001exact}
\begin{equation}
\sinh_q y = \frac{1}{2} \left(e^{y} - q e^{-y}\right),
\end{equation}

\begin{equation}
\cosh_q y = \frac{1}{2} \left(e^{y} + q e^{-y}\right), \qquad \cosh_q y = \frac{\sinh_q y}{\tanh_q y}.
\end{equation}
and
\begin{equation}
\tanh_q y = \frac{1}{\coth_q y}, \qquad \cosh_q^{2} y = q + \sinh_q^{2} y,
\end{equation}

By substituting new variables:
\begin{equation}
\nu = \sqrt{\frac{-2 \mu \mathcal{E}_n^{q}}{\alpha^{2}\hbar^{2}}}, \qquad \xi = \frac{\left(\Delta^{2}-\frac{1}{4}\right)}{\alpha^{2}},
\end{equation}
and using a new parameter $z = \tanh_q(\alpha r)$, equation (2) turns

\begin{equation}
\left\{\frac{d}{dz} \left[ (1 - z^{2}) \frac{d}{dz} \right] + \left(\frac{\xi}{q} - \frac{\nu^{2}}{1 - z^{2}} \right)\right\} \Psi(z) = 0.
\end{equation}

Substituting $\Psi(z) = (1 - z^{2})^{\frac{\nu}{2}} f(z)$
in equation (6), we obtain:

\begin{eqnarray}
(1 - z^{2}) f''(z)- 2z(\nu + 1) f' \nonumber\\+  \left[\frac{\xi}{q} + \frac{\nu(\nu + 2) z^{2}}{1 - z^{2}} - \nu+ \frac{3 \nu z^{2}}{1 - z^{2}} \right] f(z) = 0.
\end{eqnarray}

Equation (7) is in the form of the hypergeometric differential equation,
of the solution \cite{edet2020thermal, abramowitz1972handbook,arai2001exact, eugrifes1999exact}:
\begin{equation}
\Psi_n^{q} =N^{q} (1 - z^{2})^{\frac{\xi}{2}}\, {}_{2}F_{1} \left( -n,-n + 2\Tilde{w} , -n + \Tilde{w}  + \frac{1}{2};\frac{1 - z}{2}\right),
\end{equation}
where
\[{}_{2}F_{1}\left(-n,-n + 2\Tilde{w} ,-n + \Tilde{w}  + \frac{1}{2};\frac{1 - z}{2}\right)\]
are the hypergeometric polynomials of degree $n$ and
\begin{equation}
\Tilde{w}  = \sqrt{\frac{1}{4} + \frac{\xi}{q}}.
\end{equation}
The corresponding non-relativistic $q$-deformed eigenvalues read
\begin{equation}
\mathcal{E}_{n}^{q}=- \frac{\alpha^{2} \hbar^{2}}{2 \mu}\left[\left(n + \frac{1}{2}\right)- \sqrt{ \frac{1}{4} + \frac{\Delta ^2-\frac{1}{4}}{\alpha ^2 q^2}} \right]^{2}.
\end{equation}
$dE/dn = 0,$ can be used to obtain the maximum val
\begin{equation}
    n_{\text{max}} =\sqrt{\frac{\Delta ^2-\frac{1}{4}}{\alpha ^2 q^2}+\frac{1}{4}}-\frac{1}{2}
\end{equation}

\section{Quantum Otto Cycle: \texorpdfstring{$\mathcal{Q}^{q}_{\text{h}}$ and $\mathcal{Q}^{q}_{\text{c}}$}{Quantum Otto Cycle: Qh,and Qc}} \label{Ottoequations}

Here we present the input and output heat in a given cycle.
\begin{eqnarray}
 \mathcal{Q}_h^{q}= \frac{\sqrt{ \beta_\text{c} \text{p}_c}  \text{p}_{h} \left(\Upsilon_0\right)}{2 \sqrt{\pi } (\beta_\text{c} (-p_\text{c}))^{3/2} \text{erfi}\left(\sqrt{\beta_\text{c}} \sqrt{p_\text{c}} \sigma_\text{c}\right)} \nonumber\\ +\frac{\sqrt{\pi } \sqrt{\beta_\text{h} (-p_\text{h})} \text{erfc}\left(\sigma_\text{h} \sqrt{\beta_\text{h} (-p_\text{h})}\right)-2 \beta_\text{h} p_\text{h} \text{$\sigma $h} e^{\beta_\text{h} p_\text{h} \sigma_\text{ h}^2}}{2 \sqrt{\pi } \beta_\text{h}^{3/2} \sqrt{p_\text{h}} \text{erfi}\left(\sqrt{\beta_\text{h}} \sqrt{p_\text{h}} \sigma_\text{h}\right)}
 \end{eqnarray}
where $\beta_\text{c}=\frac{1}{T_\text{c}}, \beta_\text{h}=\frac{1}{\text{T}_h}$
\begin{align}
   \Upsilon_0 = \text{erf}\left(\sigma_\text{c} \sqrt{\beta_\text{c} (-p_\text{c})}\right)\sqrt{\pi }  \left(1-2 \beta_\text{c} p_\text{c} (\sigma_\text{c}-\sigma_\text{h})^2\right) + \Upsilon_1+ \Upsilon_2,\\
    \Upsilon_1 = 2 \sqrt{\beta_\text{c} (-\text{pc})} (\sigma_\text{c}-2 \sigma_\text{h}) e^{ \beta_\text{c} p_\text{c} \sigma_\text{c}^2},\\
\Upsilon_2 = \sqrt{\pi } \left(2 \text{$\beta $c} p_\text{c} (\sigma_\text{c}-\sigma_\text{h})^2-1\right),\\
\text{p}_h=\frac{\alpha_\text{h}^2 \hbar ^2}{2 \mu }, \text{p}_c=\frac{\alpha_\text{C}^2 \hbar ^2}{2 \mu },\\
\sigma_\text{h}=\frac{1}{2}-\sqrt{\frac{\Delta^2-\frac{1}{4}}{\alpha_\text{h}^2 \text{q}^2}+\frac{1}{4}}, \sigma_\text{c}=\frac{1}{2}-\sqrt{\frac{\Delta^2-\frac{1}{4}}{\alpha_\text{c}^2 \text{q}^2}+\frac{1}{4}}.
\end{align}
\begin{eqnarray}
\mathcal{Q}_c^{q} =  \frac{\sqrt{\beta_\text{h} p_\text{h}} p_\text{c}  \left( \Upsilon_{3}\right)}{2 \sqrt{\pi } (\text{$\beta $h} (-p_\text{h}))^{3/2} \text{erfi}\left(\sqrt{\text{$\beta $h}} \sqrt{p_\text{h}} \sigma_\text{h}\right)} \nonumber\\ +\frac{\sqrt{\pi } \sqrt{\text{$\beta $c} (-p_\text{c})} \text{erfc}\left(\text{$\sigma $c} \sqrt{\beta_\text{c} (-p_\text{c})}\right)-2 \beta_\text{c} \text{pc} \text{$\sigma $c} e^{ \beta_\text{c} p_\text{c} \sigma_\text{c}^2}}{2 \sqrt{\pi } \beta_\text{c}^{3/2} \sqrt{p_\text{c}} \text{erfi}\left(\sqrt{\beta_\text{c}} \sqrt{p_\text{c}} \sigma_\text{c}\right)}
\end{eqnarray}
where
\begin{align}
\Upsilon_3 = \sqrt{\pi } \text{erf}\left(\sigma_\text{ h} \sqrt{\beta_\text{h} (-p_\text{h})}\right) \left(1-2 \beta_\text{h} p_\text{h} (\sigma_\text{c}-\sigma_\text{ h})^2\right) + \Upsilon_4+ \Upsilon_5,\\
\Upsilon_4 = 2 \sqrt{ \beta_\text{h} (-p_\text{h})} (\text{$\sigma $h}-2 \sigma_\text{c}) e^{\beta_\text{h} p_\text{h} \sigma_\text{h}^2},\\
\Upsilon_5 =\sqrt{\pi } \left(2 \beta_\text{h} p_\text{h} (\sigma_\text{c}-\sigma_\text{h})^2-1\right)
\end{align}

\bibliography{References}

\end{document}